\documentclass[11pt]{JHEP3}
\usepackage{amsfonts,epsfig}

\def\be{\begin{equation}}
\def\EQ{\begin{equation}}
\def\ee{\end{equation}}
\def\EN{\end{equation}}
\def\bea{\begin{eqnarray}}
\def\beq{\begin{eqnarray}}
\def\ba{\begin{eqnarray}}
\def\eea{\end{eqnarray}}
\def\ena{\end{eqnarray}}
\def\eeq{\end{eqnarray}}
\def\ea{\end{eqnarray}}
\def\bei{\begin{itemize}}
\def\eei{\end{itemize}}
\def\bee{\begin{enumerate}}
\def\eee{\end{enumerate}}

\def\ccr{\nonumber\\}

\def\lsim{\mathrel{\rlap{\lower3pt\hbox{\hskip0pt$\sim$}}
    \raise2pt\hbox{$<$}}}         %less than or approx. symbol
\def\gsim{\mathrel{\rlap{\lower2pt\hbox{\hskip1pt$\sim$}}
    \raise2pt\hbox{$>$}}}         %greater than or approx. symbol

\def\LB{\left(}
\def\RB{\right)}
\def\lx{\left}
\def\rx{\right}

\def\a{\alpha}
\def\b{\beta}
\def\g{\gamma}
\def\l{\lambda}
\def\r{\rho}

\def\t{\tau}

\def\d{\delta}
\def\m{\mu}
\def\n{\nu}

\author{Olindo~Corradini $^{a}$ and
  Alberto Iglesias $^{b}$\\
\hskip-.25cm $^{a}$ Dipartimento  di Fisica, Universit{\`a} di Bologna
and  INFN, Sezione di Bologna\\
Via Irnerio, 46 -  Bologna I-40126, Italy
\\E-mail:
\email{corradini@bo.infn.it} \vskip2mm 
\hskip-.25cm $^{b}$ Department of Physics,
University of California,
Davis, CA 95916 U.S.A.\\ %\vskip1mm
E-mail: \email{iglesias@physics.ucdavis.edu}}

\abstract{In the present letter we find that Starobinsky's inflationary
  solution is also valid in the Dvali-Gabadadze-Porrati (DGP) model where a
  3-brane is embedded in 5-dimensional Minkowski bulk. 
  We show that such a solution is typically not supported by the 
  Self-Accelerated branch of the model, giving therefore a natural 
  selection of the conventional branch of solutions. In the absence of 
  brane induced Einstein-Hilbert term the SA branch is always selected out.
  We then study the linearized modes around all such de Sitter brane 
  solutions finding perturbative stability for a range of parameters of the 
  brane QFT.}   
 
\preprint{}

\title{Trace Anomaly Inflation in Brane Induced Gravity}

\begin{document}

%%%%%%%%%%%%%%%%%%%%%%%%%%%%%%%%%%%%%%%%%%%%%%%%%%%%%%%%%%%%%%%%%%%%%%
\section{Introduction}
%%%%%%%%%%%%%%%%%%%%%%%%%%%%%%%%%%%%%%%%%%%%%%%%%%%%%%%%%%%%%%%%%%%%%%

The DGP model of brane induced gravity~\cite{Dvali:2000hr}, a model for
modification
of gravity at large distances (cosmology at late times) provides an appealing
alternative explanation for the current accelerated expansion of the
Universe~\cite{Deffayet:2000uy,DDG}. 

In this article we study the very early Universe
in brane induced gravity. Inspired by the seminal work of
Starobinsky~\cite{Starobinsky:1980te} we consider the localized
matter as being the high energy limit of an asymptotically free QFT
(such as QCD) and search for inflationary de 
Sitter solutions (see \cite{Fabris:2000gz} for a more recent discussion about
Starobinsky's solution). A thorough study of quantum effects of brane
conformal field theories in the physics of the early (brane) 
Universe, has already been done for several other models of brane world: it includes the RS
model~\cite{Hawking:2000kj} as well as other  
variants that involve the inclusion of higher curvature bulk terms, a richer 
bulk matter content, as well as supersymmetric setups and generalizations to dS
bulks and FRW brane cosmology~\cite{Nojiri:2000eb}.

The main motivation of the present study of inflation in
brane-induced gravity models, relies on the observation
that if the DGP model successfully incorporated inflation then, interestingly
enough, it would mimic the whole 
cosmological paradigm: an initial meta-stable inflationary solution that
would decay into ``regular'' FRW cosmology. In the high energy limit, 
when the brane curvature satisfies $R > M_{\rm GUT}^2$, the brane stress tensor
is well approximated by the trace anomaly-generated 
stress tensor quadratic in brane curvatures~\cite{Vilenkin:1985md} acting as 
a source for the junction 
condition associated to the 4d modified Einstein equations of motion. 
Such a junction condition will, as usual, lead to the brane Friedman equation.

We analyze the cosmology resulting from the model, with and
without the inclusion of an induced non-conformal EH term on the brane. 
We then proceed to study the perturbative spectrum and 
find the conditions on the parameters of the brane quantum field theory such
that there is a period of inflation that is long enough.

In the original DGP model there is a branch of 
"self-accelerated" solutions~\cite{Deffayet:2000uy} in which late 
time acceleration could be driven solely by the effect of the modified 
gravitation without the need
 for any energy density. 
This branch of solutions, whose consistency with data has been extensively 
studied (see, {\it e.g.}, \cite{sab}), 
appears to be unstable both perturbatively~\cite{Luty:2003vm,CGKP} 
(see, however, the possible caveats to this conclusion~\cite{DGI}) and 
non-perturbatively (with hints in the exact solutions~\cite{K,DGPR} and 
a recent discussion in \cite{GKMP} - see also, \cite{IKPT}).
A welcome feature that results from our approach is that once the quantum 
effects that drive inflation are included the conventional
(as opposed to the self-accelerated) branch of solutions of the model is
naturally selected, if the induced Planck mass is small. On the other hand if
the induced Planck mass is large the self-accelerated branch results
perturbatively unstable.

%%%%%%%%%%%%%%%%%%%%%%%%%%%%%%%%%%%%%%%%%%%%%%%%%%%%%%%%%%%%%%%%%%%%%% 
\section{The Setup}
%%%%%%%%%%%%%%%%%%%%%%%%%%%%%%%%%%%%%%%%%%%%%%%%%%%%%%%%%%%%%%%%%%%%%%

We consider a five-dimensional action which includes the Einstein-Hilbert term 
$S_{EH}$ and a bulk matter action $S_B$ which we will assume to give rise to a 
perfect fluid type of stress tensor 
$S^M_N=-2g^{MR}\d S_B/\d g^{RN}={\rm diag}(-\rho_B,p_B,p_B,p_B,p_y)$.  
We will be interested in studying brane-cosmological bulk solutions of the form
\bea
 ds^2 = -N^2(\t,y)d\t^2 +A^2(\t,y)d\Omega_k^2 + B^2(\t,y)dy^2
\eea
where $y$ is the coordinate orthogonal to the hypersurface $\{y=0\}$ that we 
will refer to as ``the brane'' and  $d\Omega_k^2=\Omega_{mn}(x^m) dx^m dx^n$ 
is a unit radius maximally-symmetric space, characterizing the brane spatial 
section.
We will assume that some matter is confined on the brane. In particular, we
will consider the backreaction caused by the renormalized stress tensor of a
set of massless conformal 
quantum fields on an isotropic homogeneous (brane) universe, along the lines of
what originally done by Starobinsky~\cite{Starobinsky:1980te}. Such massless
quantum fields are to be understood as the high energy limit of the (massive)
fields of an asymptotically free QFT.  The result will
be twofold: on the one hand we generalize Starobinsky's idea to the DGP brane
world model. On the other hand we study brane-induced gravity (DGP
model~\cite{Dvali:2000hr}) in case 
when the matter stress tensor on the brane is dominated by the
trace-anomaly-induced one.

It was
shown~\cite{Binetruy:1999hy} that, assuming
there is no bulk-brane energy exchange, the $\t\t$ and $yy$ bulk equations of 
motion take the simple form~\footnote{We denote $\dot{} =
  \partial_\t$ and $'=\partial_y$.} 
\bea
F' &=& -{2\kappa^2\over 3} A' A^3\rho_B \label{F'}\\
\dot F &=& {2\kappa^2\over 3} \dot A A^3 p_y \label{F.} 
\eea
where 
\bea
F(t,y)=\left({A' A\over B}\right)^2-\left({\dot A A\over N}\right)^2-k A^2~,
\eea
whereas the $\t y$ equation of motion reads
\bea
\frac{N'}{N}\frac{\dot A}{A}-\frac{\dot A'}{A} = -\frac{N}{A}\left(\frac{\dot
  A}{N}\right)' =0~.
\label{ty}
\eea
Equation~(\ref{F'}) can be integrated and yields
\bea
F = -{\kappa^2\over 6}\rho_B A^4 -C~,
\eea
where $C$ is an integration constant that was found~\cite{Shiromizu:1999wj} to
be related to the bulk Weyl tensor. Hence,
\bea
\left( {A' \over AB}\right)^2 =\left( {\dot A \over AN}\right)^2 +{k\over
  A^2}-{C\over A^4}-{\kappa^2\over 6}\rho_B~.    
\label{1-int}
\eea
We will be interested in the case of an empty, Minkowski bulk for which
$\rho_B =C=0$ so that $F=0$ and the leftover Einstein's equation is
also automatically satisfied and (\ref{1-int}) reduces to
\bea
\left({A'\over AB}\right)^2 =\left({\dot A\over AN}\right)^2 +{k\over A^2}~.
\eea
Hence,
\bea
{A'\over AB}=-\epsilon\sqrt{\left({\dot A\over AN}\right)^2+{k\over A^2}}~,
\label{first-bulk}
\eea
which projected upon the brane becomes
\bea
{a'\over ab}=-\epsilon\sqrt{\left({\dot a\over a}\right)^2+{k\over a^2}}~,
\label{first}    
\eea
where
\bea
a(\t) &=& A(\t,0)~,\\ 
b(\t) &=& B(\t,0)~,\\
n(\t) &=& N(\t,0) \equiv 1 ~.
\eea
The l.h.s. of~(\ref{first}) is related to the extrinsic curvature on the brane
that is defined by
\bea
K_{\mu\nu} = \left. e_\mu^M e_\nu^N \nabla_M n_N\right|_{\Sigma}~,
\eea
where 
\bea
e_\mu^M &=& {\partial X^M\over \partial x^\mu} =\delta_\mu^M~,\\
e_\mu^M n_M &=& 0~,
\eea
are respectively the tangent vectors characterizing the brane embedding in the
bulk (we choose a static gauge) and the (unit) normal vector to the brane 
\bea
n_M=(0,{\bf 0},B)~,
\eea
so that 
\bea
K^0{}_0 &=& {n'\over b}~,\\
K^m{}_n&=&{a'\over ab}\delta^m{}_n=-\epsilon\sqrt{\lx({\dot a\over a}\rx)^2
+{k\over a^2}}\delta^m{}_n~, 
\quad \epsilon =\pm 1~,
\label{Kmn}
\eea
is the expression for the extrinsic curvature tensor, where we have made
use of~(\ref{first}).  The Israel junction condition links the latter to the
stress tensor of the matter localized on the brane
\bea
K_{\mu\nu} -\gamma_{\mu\nu} K = -{1\over 2 M^3} T_{\mu\nu}~,
\eea
where the lhs is evaluated at $y=0+$ and $\gamma_{\mu\nu}$ is the induced
metric on the brane, namely,  
\bea
d\tilde s^2=-d\t^2+a(\t)^2d\Omega_k^2\equiv\g_{\m\n}(x) dx^\m dx^\n~.
\eea 

Assuming the brane matter to be a
perfect fluid, from the spatial components of the extrinsic curvature we get
\bea
\epsilon\sqrt{\lx({\dot a\over a}\rx)^2+{k\over a^2}}={1\over 6 M^3} \rho~.   
\label{ED}
\eea 
If the localized matter is conformal the (v.e.v. of the) energy momentum tensor
reduces to the one generated by the trace anomaly~\cite{Starobinsky:1980te,
Davies:1977ze,Hawking:2000bb} 
\bea\label{tmn}
T_{\mu\nu}^A &=&\tilde \a \lx(-R_\m{}^\sigma R_{\n\sigma}+{2\over 3}R R_{\m\n} 
+{1\over 2} g_{\m\n} R_{\a\b}^2-{1\over 4}g_{\m\n}R^2\rx)\ccr
&&+ {\tilde \b\over 6}\lx(-2\nabla_\m\nabla_\n R+2 g_{\m\n}\Box R +2R R_{\m\n}
-{1\over 2}g_{\m\n}R^2\rx)\equiv \tilde\a\Theta^1_{\mu\nu}
+\tilde\b\Theta^2_{\mu\nu}  ~,
\eea
where $\tilde \a=k_2/6!(2\pi)^2$ and $\tilde \b=k_3/6!(2\pi)^2$ and all
tensors are calculated according to the induced metric, and yield 
%~\footnote{What
 % appears on the r.h.s. of~(\ref{friedman}) is $\rho/6$.} 
\bea
&&\epsilon \sqrt{\lx({\dot a\over a}\rx)^2 +{k\over a^2}} =
{1\over 2M^3}\lx\{\tilde\a\lx(\lx({\dot a\over a}\rx)^2+{k\over a^2}\rx)^2\rx.
\ccr
&&\lx. +\tilde\b\lx[-\lx({\ddot a\over a}\rx)^2+2{\dot a\over
    a}\lx({\ddot a\over a}\rx)^\centerdot+2{\dot a\over a} \lx(\lx({\dot a\over
    a}\rx)^2 +{k\over a^2}\rx)^\centerdot+\lx(\lx({\dot a\over a}\rx)^2
  +{k\over a^2}\rx)^2\rx]\rx\} ~,
\label{friedman}
\eea
that is the brane Friedman equation which can be written as  
\bea
\epsilon\ v^{1/2} a^2 = {1\over 2 M^3}\Biggl[ \tilde\a {v^2\over a^4}
+ \tilde\b\left({vv_{aa}\over a^2}-2{vv_a\over a^3}-{v_a^2\over
  4a^2}-kv_{aa}+3{kv_a\over a}\right)\Biggr]~,
\eea
by means of the transformation $v(a)=(a\dot a)^2 +k a^2$. One possible set of 
solutions of the latter equation can be achieved by setting $v = H^2 a^4$ 
(with $H^2$ constant),
upon which the term in the round parenthesis vanishes (lower ``indices'' $a$
denote derivatives w.r.t. $a$); the leftover terms
uniquely fix the constant,   
\bea
&& (a\dot a)^2 +k a^2 = H^2 a^4~,\ccr
&& \tilde \a H^3  = 2M^3\epsilon~, 
\label{sol:dS}
\eea
and select the allowed branch, that is $\epsilon=+1$ for $k_2>0$ and
$\epsilon=-1$ in the opposite case. In other words de Sitter type of solution 
\bea
a(\t)=\left\{
\begin{array}{ll}
 a_0\ \exp({Ht}),& \quad k=0\\
 H^{-1}\cosh(Ht),& \quad k=+1\\
 H^{-1}\sinh(Ht),& \quad k=-1
\end{array}\right.
\label{dS}
\eea
holds in our brane world setup. However, since~\footnote{$N_S$ is the
  number of real scalar fields, $N_F$ the number of Dirac fermions and $N_V$
  the number of vector fields~\cite{Duff:1977ay}.}  
\bea\label{k2}
k_2 = N_S + 11 N_F +62 N_V >0~,
\eea
only the conventional branch ($\epsilon =+1$) is allowed for such a solution. 
In fact, a negative contribution to $k_2$ can be obtained from quantum fields
that have a {\em wrong} sign in the kinetic term, in other words {\em  ghosts}:
since the trace anomaly in four dimensions comes from a
triangle diagram describing the correlation function of three stress tensors,
changing the sign of the action for the quantum fields changes that of the
stress tensor and in turn that of the anomaly. The previous argument clearly 
raises an
issue concerning the (quantum) stability of the branch $\epsilon =-1$, in the
present setup. Such a branch corresponds to the so-called self-accelerated
branch of the original DGP model \cite{Deffayet:2000uy}.~\footnote{Let
us stress that here we refer to $\epsilon=+1\ (\epsilon=-1)$ as the
conventional (self-accelerated) branch just to make contact with
analogous definitions given in DGP cosmology~\cite{Deffayet:2000uy}
where the classification is related to the sign of the
total energy density as in~(\ref{ED}): our reference to such names is
just nomenclature and is not linked to the properties of the
solutions. In fact, in our case both dS branches are, strictly
speaking, self-accelerated meaning that the acceleration is
geometrical and not driven by sources. For the sake of clarity in
fact, the ``true'' conventional branch
would be flat 4d Minkowski that is obviously a solution of~(\ref{ED}).} 
Hence, the argument we gave above would a priori seem to confirm the more 
pessimistic points of view. Note however that such an argument does not imply 
that the self-accelerated branch is altogether excluded by (\ref{friedman}); in
fact the second term on r.h.s.~of (\ref{friedman}) is not positive definite - 
it is related to the trivial anomaly - and thus it might allow for solutions 
even for $\epsilon =-1$.  

Before concluding this section let us also point out that in the above solution
the inflation rate $H$ is set in by the fundamental scale $M$ and depends upon
$4d$ physics only through the numerical combination $k_2$. In the (DGP) brane
world scenario  the fundamental scale is taken to be much smaller that the 4d
Planck mass: therefore the curvature scale $R\sim H^2$ is also much smaller
than $M_P^2$ and thus lower than $M_{\rm GUT}^2$, so that the validity of the
approximation~(\ref{tmn}) appears not to be guaranteed. However, we will see
in the next section that the inclusion of an induced non-conformal
Einstein-Hilbert term on the brane enhances the value of the de Sitter scale.
A brane-induced E-H term is also a necessary ingredient in order to recover 4d
gravity between brane massive sources.

%%%%%%%%%%%%%%%%%%%%%%%%%%%%%%%%%%%%%%%%%%%%%%%%%%%%%%%%%%%%%%%%%%%%%%
\subsection{Adding a non-conformal Einstein-Hilbert term on the brane}
%%%%%%%%%%%%%%%%%%%%%%%%%%%%%%%%%%%%%%%%%%%%%%%%%%%%%%%%%%%%%%%%%%%%%%

In this subsection we consider the inclusion of a non-conformal induced EH
term on the brane. Such an induced term is part of the effective action of
matter fields coupled to gravity~\cite{Dvali:2000hr}. However it was shown
that a brane EH term can be produced also at the classical level on a
tensionful brane if the bulk action includes higher curvature
terms~\cite{Corradini:2001qv}.

The presence of an induced E-H term on
the brane does not necessarily mean that the inflationary phase is no longer 
there. In fact as we now show we can have a steeper inflationary phase whose 
scale is set in by a combination of both $M$ and induced Planck mass $M_P$. 
Note in fact that the anomaly generated stress tensor, ``ameliorated'' with 
the localized Einstein tensor coming from the induced EH term, is a good
approximation for ``large'' brane curvature, $M_P^2 > R > M_{\rm GUT}^2$. The
first inequivalence guarantees that brane loop corrections are under control. 
Note also that, although bulk loops are controlled by the fundamental scale
$M$, the expansion is under control there as well since the bulk is (a patch 
of) flat Minkowski space. The aforementioned inflationary phase could be
unstable as pointed out by Starobinsky and Vilenkin~\cite{Starobinsky:1980te,
Vilenkin:1985md} and the universe would thus finally decay to ``regular'' FRW 
cosmology. We discuss about stability in the next sections.    

In this section we thus consider the inclusion of an Einstein-Hilbert term on 
the brane
\bea
S_4 = \frac{M_P^2}2 \int d^4x \sqrt{-g} R_4
\eea 
so that (\ref{sol:dS}) becomes
\bea
\epsilon m_c + H = {\tilde\a H^3\over M_P^2}
\label{sol:dSR}
\eea
with $m_c = 2 M^3/M_P^2$ being the crossover scale of 
the DGP model. Denoting $z = H/ M\lambda $, with 
$\lambda =(2\tilde\a)^{-1/3}$, equation~(\ref{sol:dSR}) can be rewritten as a 
cubic equation
\bea
z^3-2\lambda\lx({M_P\over M}\rx)^2 z-4\epsilon=0
\eea
that is of the form that allows for direct Cardano's solution
\bea
{H\over M\l}=
\lx[2\epsilon+\sqrt{4-\lx({2\l\over3}\lx({M_P\over M}\rx)^2\rx)^3}\rx]^{1/3}
+\lx[2\epsilon-\sqrt{4-\lx({2\l\over3}\lx({M_P\over M}\rx)^2\rx)^3}\rx]^{1/3}~.
\eea 

For $\epsilon=+1$ there is one real
positive solution regardless of the value of the quantity
$\l\lx({M_P/M}\rx)^2 $. 

For $\epsilon=-1$ and $0<\l\lx({M_P/M}\rx)^2 <3\cdot 2^{-1/3}$, there 
is no positive solution whereas for $\l\lx({M_P/M}\rx)^2 > 3\cdot 2^{-1/3}$ 
there are two positive solutions. 
Therefore, if the induced Planck mass is large enough, 
both branches would seem to allow expanding de Sitter solutions~(\ref{dS}). In 
fact, in the limit where $M_P \gg M$ one recovers Starobinsky's expansion rate 
$H_S\sim M_P/\sqrt{\tilde\alpha}$. Note also that the expansion rate for the
conventional (self-accelerated) branch corresponds to ${\tilde\a H^2\over
  M_P^2} > 1\ \LB{\tilde\a H^2\over M_P^2}<1\RB$. In other words a phenomenological
selection ``criterion'' that, a priori, seems to pick the conventional branch is the
highest dS expansion rate.  However, in such a case higher order
corrections (in $H/\Lambda$) to the quantum stress tensor
are not guaranteed to be under 
control, since the induced Plack mass is related to the UV cutoff of
the localized theory as $M^2_P \sim \Lambda^2 N$, with $N$ being
(roughly) the total number of fields in the theory $(N\lsim k_2)$
and thus $H_S \gsim \Lambda$.
%~\footnote{One may worry about higher loop
%corrections as well since  the trace anomaly, as opposed to chiral
%anomalies, is not one loop exact. Note however that in an
%asymptotically free theory such corrections are negligible in the UV limit.}  
An hypothetical way out to such a problem
might be the inclusion of bulk higher curvature terms that could allow
to lower the dS scale. Let us also remind the reader that, in order
for quantum gravitational corrections to be under control, $H_S < M_P$, the total number
of fields must be extremely large such that $k_2 > 6!$~\footnote{A
  severe parametrical constraint comes from the observational limit on the amplitude
of long-wavelength gravitational waves, requiring $k_3 \gg
k_2$ \cite{Vilenkin:1985md}, that appears quite unnatural. However
notice that $k_3$ is the coefficient of the trivial part of the anomaly and is
thus arbitrary.}

%%%%%%%%%%%%%%%%%%%%%%%%%%%%%%%%%%%%%%%%%%%%%%%%%%%%%%%%%%%%%%%%%%%%%%
\section{Perturbations of the de Sitter brane solution}
%%%%%%%%%%%%%%%%%%%%%%%%%%%%%%%%%%%%%%%%%%%%%%%%%%%%%%%%%%%%%%%%%%%%%%

In this section we consider small perturbations around the de Sitter brane
solution introduced previously. Let us first point out
that the stress tensor $\Theta^1_{\mu\nu}$, as defined in~(\ref{tmn}), is
not conserved in a generic background and should be supplemented by a
term $R^{\a\b} C_{\m\a\n\b}$, with $C_{\m\a\n\b}$ being the Weyl tensor. However,
linear perturbations of such a term around a maximally-symmetric
background, vanish: we can thus safely omit the latter in what follows. 

 The bulk equations and junction conditions are given by: 
\bea\label{bu}
R_{MN}&=&0~,\\
G_{\mu\nu}-m_c\left(K_{\mu\nu}-g_{\mu\nu}K\right)&=&
{T_{\mu\nu}^A\over M_P^2}+{T_{\mu\nu}\over M_P^2}~,\label{ju}
\eea
where we added to the r.h.s.~of (\ref{ju}) an extra subleading $T_{\mu\nu}$ 
component.

We want to solve for the perturbations around the $k=0$ de Sitter background 
solution (see the appendix) in a convenient gauge, namely,
\bea
g_{\m\n}=N^2(y)\g_{\m\n} +\tilde h_{\m\n}~,~~~~~ 
g_{yy}=1~~~~~g_{\m y}=0~, ~~~~~
\nabla^\m \tilde h_{\m\n}=\nabla_\n \tilde h~. 
\eea
where $N(y)=1-\epsilon H|y|$, $\g_{\m\n}$ has curvatures 
$R_{\m\n}=3H^2\g_{\m\n}$ and $R=12H^2$ and $\epsilon=\pm 1$ 
($\epsilon=+1$ is the conventional branch: $|y|\le H^{-1}$).  

Let us first gather some results:
\bea
\d R_{\m\n}&=&{1\over 2}\nabla_\m\nabla_\n\tilde h
-{1\over 2}\Box\tilde h_{\m\n}+4H^2\tilde h_{\m\n}-H^2\g_{\m\n}\tilde h~,\\
\d R&=&-3 H^2 \tilde h~,
\eea
Thus,
\bea
\d G_{\m\n}=-{1\over 2}\left(\Box\tilde h_{\m\n}-
\nabla_\m\nabla_\n\tilde h\right)-2H^2\tilde h_{\m\n}+
{1\over 2}H^2\g_{\m\n}\tilde h~,
\eea
From the $yy$ equation we get $\g^{\m\n}\d K_{\m\n}=-\epsilon H \tilde h/2$ 
\cite{DGI} (cfr. eqs. (15-16)), which combined with the background
equation~(\ref{sol:dSR}) gives  
\bea
\d \left(-m_c (K_{\m\n}-g_{\m\n}K)\right)=
-{1\over 2}m_c \partial_y\tilde h_{\m\n}
-\frac{1-\a}2 H^2\g_{\m\n}\tilde h+4(1-\a)H^2\tilde h_{\m\n}~,
\eea
where $\a=\tilde\a H^2/M_P^2$ and, for future reference, 
$\b=\tilde \b H^2/M_P^2$.
Therefore, the junction conditions (\ref{ju}) and its trace yield:
\bea\label{ju2}
&&-{1\over 2}\LB m_c\partial_y+(1-2\a-4\b)\Box-4H^2(1-{3\over 2}\a-2\b)\RB
\tilde h_{\m\n}=\nonumber\\
&&~~~{1\over M_P^2}\LB T_{\m\n}-{1\over 3}\g_{\m\n}T\RB
-{1\over2}\LB(1-2\a-6\b)\nabla_\m\nabla_\n-(1-2\a-2\b)H^2\g_{\m\n}\RB\tilde h~,
\\ &&\LB -{1\over2}\epsilon H m_c-\a H^2+\b \Box \RB\tilde h={T\over3M_P^2}~.
\label{tr2}
\eea 
Note that for non vanishing $\b$ the trace $\tilde h$ becomes a propagating 
degree of freedom.

A way towards the solution of this system is through the shift to transverse 
traceless metric perturbations $h_{\m\n}$ by introducing the brane bending mode
 $\varphi(x)$ \cite{Garriga:1999yh}
\bea\label{sh}
\tilde h_{\m\n}=h_{\m\n}+{2N\over H}\Pi^+_{\m\n}\varphi~,
\eea
with $\Pi^\pm_{\m\n}=\nabla_\m\nabla_\m\pm\g_{\m\n}H^2$.

The system (\ref{ju2}), (\ref{tr2}) and the bulk $\m\n$ equations becomes:
\bea\label{ju3}
&&
-{1\over2}\LB m_c\partial_y+(1-2\a-4\b)\Box-4H^2(1-{3\over2}\a-2\b)\RB h_{\m\n}
=\nonumber\\ 
&&~~~
{1\over M_P^2}\LB T_{\m\n}-{1\over 3}\g_{\m\n}T\RB
+\LB -\epsilon m_c-2H(\a+4\b)\RB\Pi^+_{\m\n}\varphi
-2{\b\over H}\Pi^+_{\m\n}{\cal Q}\varphi~,\\ 
\label{tr3}&&
\LB\epsilon H m_c+2\a H^2-2\b\Box\RB{\cal Q}\varphi={HT\over3M_P^2}~,\\
&&\LB N^2\partial_y^2-{\cal O}\RB h_{\m\n}=0~,\label{bu3}
\eea
where we use
\bea
{\cal O}&=&4H^2-\Box~,\\
{\cal Q}&=&-4H^2-\Box~,
\eea
and the identity
\bea
{\cal O}\Pi^+_{\m\n}\phi=\Pi^-_{\m\n}{\cal Q}\phi~.
\eea
valid for any scalar $\phi$.
Combining (\ref{ju3}) and (\ref{bu3}) into a single equation gives:
\bea
&&\left[ m_c\partial_y^2-\LB {m_c\over N^2}+ 2\d(y)\RB{\cal O}
-2\d(y)\LB(2\a+4\b)\Box-2H^2(3\a+4\b)\RB\right] h_{\m\n}
=\nonumber\\ 
&&
-4\d (y)\left[{1\over M_P^2}\LB T_{\m\n}-{1\over 3}\g_{\m\n}T\RB
-\LB \epsilon m_c+2H(\a+4\b)\RB\Pi^+_{\m\n}\varphi
-2{\b\over H}\Pi^+_{\m\n}{\cal Q}\varphi\right]~,
\eea
which by making use of the condition (\ref{sol:dSR}) derived from the 
background equations, {\it  i.e.}, $\epsilon m_c=(\a-1)H$, can be solved 
to yield 
\bea\label{sol}
\!\!{1\over2}h_{\m\n}=S({\cal O})\LB {1\over M_P^2}\LB T_{\m\n}
-{1\over3}\g_{\m\n}T
\RB-\LB\epsilon m_c+2H(\a+4\b)\RB\Pi^+_{\m\n}\varphi
-2{\b\over H}\Pi^+_{\m\n}{\cal Q}\varphi\RB
\eea
where,
\bea\label{s}
S(z)&=&{N^{q(z)}\over (1-\a)\LB z-H^2q(z)\RB-(\a+4\b)\LB z-2H^2\RB}~,\\
q(z)&=&{1\over2}\LB 1+\epsilon\sqrt{1+4{z\over H^2}}\RB~.\label{q}
\eea
We recover $\tilde h_{\m\n}$ via (\ref{sh}) with $\varphi$ given by the 
solution of (\ref{tr3}).

Notice that we have correlated the two possible solutions (\ref{q}) with the 
two branches of the model via the $\epsilon$ parameter. This implies a choice 
of boundary conditions at $y=\infty$: had these been uncorrelated, 
non-normalizable modes would be contributing to the amplitudes implying the 
presence of sources far away into the bulk. 

Having the solution for the perturbations (\ref{sol}) we now concentrate on the
one graviton exchange amplitude between brane localized sources $T$ and 
$T^\prime$ (we follow the approach presented in \cite{DGI}; see \cite{G} for a 
pedagogical introduction). To this end,
it is useful to decompose the corresponding conserved energy momentum tensors 
$T_{\m\n}$ and $T^{\prime \m\n}$ into transverse traceless, longitudinal and 
trace parts:
\bea
T_{\m\n}=T^{TT}_{\m\n}+{1\over4}\g_{\m\n}T+{1\over3}P_{\m\n}{1\over{\cal Q}}T~.
\eea 
where $P_{\m\n}=\nabla_\m\nabla_\n-{1\over4}\g_{\m\n}\Box$ (for which the 
identity $P_{\m\n}f({\cal Q})\phi=f({\cal O})P_{\m\n}\phi$ holds for any 
smooth $f$ and scalar $\phi$).

We obtain the following expressions:
\bea
{1\over 2}\tilde h_{\m\n}&=&{1\over M_P^2}S({\cal O})T^{TT}_{\m\n}
+{N\over H}\Pi^+_{\m\n}\varphi~,\\
\varphi&=&-{1\over3\xi H M_P^2}\LB{1\over{\cal Q}}-{2\b\over2\b{\cal Q}-\xi
  H^2}\RB T~,
\label{bb}
\eea
where $\xi=1-3\a-8\b$.

Next, let us introduce the Lichnerowicz operator  $\Delta$ with
the following properties
\bea
(\Delta -4H^2)T_{\m\n}^{TT}&=&{\cal O}T_{\m\n}^{TT}~,\\
(\Delta -4H^2)\gamma_{\mu\nu}\phi&=&\gamma_{\mu\nu}{\cal Q}\phi~,\\
(\Delta -4H^2)P_{\mu\nu}\phi&=&P_{\mu\nu}{\cal Q}\phi~.
\eea
This gives,
\bea
S({\cal O})T^{TT}_{\m\n}&=&S(\Delta-4H^2)T_{\m\n}
-{1\over 4}S({\cal Q})\g_{\m\n}T
-{1\over3}\nabla_\m\nabla_\n {S({\cal Q})\over {\cal Q}}T
+{1\over 12}\g_{\m\n}{\Box\over{\cal Q}}S({\cal Q})T~,\nonumber\\
T^{\prime\m\n}S({\cal O})T^{TT}_{\m\n}&=&T^{\prime\m\n}S(\Delta-4H^2)T_{\m\n}
+{1\over12}T^\prime\LB{\Box\over{\cal Q}}-3\RB S({\cal Q})T+\nabla_\m(\cdots)~.
\eea

We compute the amplitude,
\bea
{\cal A}&=&{1\over2}\int d^4x \sqrt{-\gamma}\ T^{\prime\m\n}\tilde h_{\m\n}~,
\eea
obtaining:
\bea
{\cal A}&=&
{1\over M_P^2}\int d^4x \sqrt{-\gamma}\ \Biggl\{
T^{\prime\m\n}S({\Delta-4H^2})T_{\m\n}
\nonumber\\
&-&{1\over3}T^\prime\left[\LB 1+{H^2\over{\cal Q}}\RB S({\cal Q})
+{N\over\xi{\cal Q}}-{2\b N\over\xi\LB2\b{\cal Q}-\xi H^2\RB}\right]T
\Biggr\}~.
\eea

Let us focus on the conventional branch ($\epsilon=+1$). In this case, there 
could be poles at ${\cal Q}=2H^2$ (vanishing denominator of $S({\cal Q})$), 
at ${\cal Q}=0$ and at $2\b{\cal Q}=\xi H^2$. We will also discuss below the 
interpretation of the other zero of the denominator of $S({\cal Q})$.

\bigskip

The ${\cal Q}=2H^2$ ($\Box=-6H^2\equiv -2\Lambda$) pole, together with the one 
at $\Delta=6H^2$ corresponds to a massless spin-2 state on the de Sitter 
background (with the correct tensorial structure).

The would be ${\cal Q}=0$ ($\Box=-4H^2$) pole has vanishing residue 
(at that point $H^2S(0)+N/\xi=0$).

\bigskip

To go further into the analysis, let us find another expression for the 
amplitude. We define a complex variable $w\equiv-z-H^2/4$ such that 
$\tilde S(w)=S(-w-H^2/4)$ has a branch cut in the positive real $w$ axis.
We get the alternative form 
\bea
\tilde S(w)=\frac1{2\pi i}\oint_{\Gamma} \frac{ds}{s-w} \tilde S(s)
-\frac{1}{w_0-w}\lim_{s\rightarrow w_0}(s-w_0)\tilde S(s)
\eea 
via a contour integral $\Gamma$ which goes 
around the cut just below and above the positive $w$ axis and closes at 
infinity through a counterclockwise circle. The residue of the pole in the 
first Riemann sheet, $w_0$, plus the jump in the imaginary part across the cut give   
\bea\label{cont}
\tilde S(w)&=&{c\over w_0-w}
+\int_0^\infty {ds\over s-w}\r(s)~,\\
c&=&{3\over2-5\a-12\b}~,\\
\r(s)&=&{1\over \pi}{(\alpha-1)H\sqrt{s}\over (1-\a)^2H^2s+\LB(1-2\a-4\b)s
+3H^2({1\over4}-\a-3\b)\RB^2}~.\label{sd}
\eea
or equivalently,
\bea\label{spec}
S({\cal Q}\equiv-\Box-4H^2)=
{c\over -\Box -2\Lambda}+\int^\infty_{{3\over2}H}
dm {2m\over m^2-\Box-2\Lambda}~\r\LB m^2-9H^2/4\RB~.
\eea
The simple pole at $w=w_0\equiv -9H^2/4$ in (\ref{cont}) corresponds to the 
massless state ($\Box+2\Lambda=0$ in (\ref{spec}), see equivalence with 
expressions in \cite{Porrati:2000cp}) while the integral has 
contributions from the continuum of massive KK modes 
$0\le w<\infty$, {\it i.e.}, $9H^2/4\le \Box+2\Lambda<\infty$ with positive 
definite spectral density (\ref{sd}) as long as $\alpha>1$
(a resonance representing the other zero in the denominator 
of $S$ - see (\ref{s})). With these conventions, the massless state has 
positive norm as long as $c$ is positive, which for the regime of interest
for us, namely, $\a\gsim 1$, implies $\b\lsim -1/4$.  

\bigskip

The $2\b{\cal Q}=\xi H^2$ pole has positive residue (corresponding to an extra 
scalar state with positive norm) as long as $\xi$ is positive, which for 
$\a\gsim 1$ coincides with the positivity of norm of the massless state. 
The squared mass (eigenvalue of $\Box+2H^2$) of this extra state is 
\beq\label{mh}
m_h^2=-2H^2-{\xi\over 2 \b} H^2~.
\eeq  
By recalling the relation to the parameters $k_2$ (\ref{k2}) and $k_3$ in 
\cite{Starobinsky:1980te}
\bea\label{xi}
\xi&=&1-{1\over 2880\pi^2}(3k_2+8k_3)\LB {H\over M_P}\RB^2~,\\
\b&=&{k_3\over 2880\pi^2}\LB {H\over M_P}\RB^2~,
\eea 
we see that a modest negative $k_3$ partially compensating $k_2$ in (\ref{xi}) 
leads to a theory free of ghost-like instabilities around a de Sitter 
background. Furthermore,
in the original Starobinsky setup a future instability is 
found~\cite{Starobinsky:1980te, Vilenkin:1985md} due to a light tachyon. 
This slow instability which could be 
responsible for the exit from a long inflationary phase 
persists in our setup. Keeping $\xi>0$ and focusing in the regime 
$\alpha\gsim 1$, $\beta\lsim -1/4$ gives a tachyonic 
mass $m_h^2\gsim -2H^2$ to the extra scalar state.

On the self-accelerated branch ($\epsilon =-1$) the pole $w=w_0$ sits in the
first Riemann sheet provided one chooses the opposite prescription for
$\sqrt{-w}$ \cite{DGI}. With this choice we obtain that the spectral density has
the same form as in~(\ref{sd}) though now is negative definite, since $\alpha
<1$: it can be interpreted as due to a tower of negative-norm KK states.    

\section{Conclusions}

We considered the high energy limit of asymptotically-free brane matter in 
Minkowski bulk and studied Starobinsky's trace-anomaly driven inflationary 
solutions.
We studied perturbations about such de Sitter solutions and found
that the system could display no instabilities. 
In particular, in the conventional branch there is a range of 
parameters (dictated by the number of scalar, vector and fermionic components) 
of the brane QFT such that there are no ghost-like excitations but a slow 
instability to end the inflationary phase. 

In absence of any induced
brane EH term, we found that only the {\it conventional} branch allows 
the aforementioned solution, becoming naturally selected. 
In the presence of an induced EH term the phenomenon is more complex: 
for small induced Planck mass, 
$ M_P < M \tilde\alpha^{1/6}$, only the conventional branch is allowed,
whereas for big induced Planck mass $ M_P >  M \tilde\alpha^{1/6}$ both
branches are allowed. In particular in the limit $M_P\gg M$ both branches
converge to the Starobinsky's solution. For the self-accelerated branch 
we have $H\lsim H_S$ that corresponds to $\alpha \lsim 1$ ($\alpha=1$ being
Starobinsky's solution),
whereas for the conventional branch we have $H \gsim H_S$ and
$\alpha \gsim 1$.  However, the self-accelerated branch turns out to be
unstable regardless of the value of $M_P$.

\acknowledgments{It is a pleasure to thank F.~Bastianelli, G.~Gabadadze, 
N.~Kaloper, S.~Odintsov and L.~Sorbo for very useful discussions and
the referee for very helpful criticisms. We are grateful to the 
organizers of the Simons Workshop 2006 held at SUNY at Stony Brook, where this 
work was initiated, for their hospitality. AI was supported by DOE Grant 
DE-FG03-91ER40674. OC is grateful to the Department of Physics of UC Davis 
and to the Institute of Cosmology and Gravitation of the University of 
Portsmouth for support and hospitality while parts of this work were completed.
} 

%\newpage
\appendix
\section{Bulk solutions}
We briefly sketch the derivation of the bulk solutions corresponding to the
dS cosmologies studied above. As shown
by~\cite{Binetruy:1999hy,Deffayet:2000uy} the assumption of zero 
bulk-brane energy exchange (cfr. eq. (\ref{ty})) allows to write
\bea
\frac{\dot A}{N}= \alpha(\t)~.
\eea
Thus setting the brane ``clock'' such that $N(\t,0) =1$, and using the bulk
equation of motion~(\ref{first-bulk}) we get 
\bea
A(\t,y) &=& a(\t) -\epsilon |y| \sqrt{k+\dot a^2}\\
N(\t,y) &=& 1- \epsilon |y| \frac{\ddot a}{\sqrt{k+\dot a^2}}
\eea
and therefore, using the explicit solutions described in the main part of the
paper we obtain
\bea
A(\t ,y) &=& a(\t)\ (1-\epsilon H |y|)\\
N(y) &=& 1-\epsilon H|y|
\eea
and thus
\bea
ds^2 = N(y)^2 d\tilde s^2 + dy^2 = N(y)^2 \biggl(-d\t^2
+a^2(\t)\, d\Omega_k^2\biggr)+dy^2 ~,
\eea
with $|y|< H^{-1}$ for the conventional branch $\epsilon=+1$ and $|y|< \infty$
for the self-accelerated branch $\epsilon=-1$. In conformal coordinated the
latter reads
\bea
ds^2 = e^{-\epsilon H |z|}\biggl(-d\t^2 +a^2(\t)\, d\Omega_k^2  +dz^2\biggr)
\eea
with $z\in {\mathbb R}$.

%\newpage

\end{document}